\documentclass[%
twocolumn,superscriptaddress,
 amsmath,amssymb,
 aps,
pra,
]{revtex4}

\usepackage{graphicx}
\usepackage{dcolumn}
\usepackage{bm}
\usepackage{epstopdf}
\usepackage{graphicx}
\usepackage{subfigure}

\usepackage[colorlinks,
linkcolor=blue,
anchorcolor=blue,
urlcolor=blue,
citecolor=blue]{hyperref}

\usepackage[mathlines]{lineno}

\newcommand{\rr} {\boldsymbol{r}}

\begin{document}

\title{ Small amplitude collective modes of a finite-size unitary Fermi gas in deformed traps }

\author{Na Fei}
\affiliation{State Key Laboratory of Nuclear
Physics and Technology, School of Physics, Peking University,  Beijing 100871, China}

\author{J.C. Pei}
\email{peij@pku.edu.cn}
\affiliation{State Key Laboratory of Nuclear
Physics and Technology, School of Physics, Peking University,  Beijing 100871, China}

\author{K. Wang}
\affiliation{State Key Laboratory of Nuclear
Physics and Technology, School of Physics, Peking University,  Beijing 100871, China}

\author{M. Kortelainen}
\affiliation{Department of Physics, P.O. Box 35 (YFL), University of Jyvaskyla, FI-40014 Jyvaskyla, Finland}
\affiliation{Helsinki Institute of Physics, P.O. Box 64, FI-00014 University of Helsinki, Finland}

\begin{abstract}

We have investigated collective breathing modes of a unitary Fermi gas in deformed harmonic traps.
The ground state is studied by the Superfluid Local Density Approximation (SLDA) and small-amplitude collective modes are studied by the iterative Quasiparticle Random Phase Approximation (QRPA).
The results illustrate the evolutions of collective modes of a small system in traps from spherical to elongated or  pancake deformations.
For small spherical systems, the influences of different SLDA parameters are significant, and, in particular, a large pairing strength can shift up the oscillation frequency of collective mode.
The transition currents from QRPA show that the compressional flow patterns are nontrivial and dependent on the deformation.
Finally, the finite size effects are demonstrated to be reasonable when progressing towards larger systems.  
Our studies indicate that experiments on small and medium systems are
valuable for understanding effective interactions in systems with varying sizes and trap deformations.

\end{abstract}

\maketitle


\section{\label{sec:level1}INTRODUCTION}

The studies of strongly interacting ultracold atomic gases have interdisciplinary interests in quantum many-body systems~\cite{strinati,zwer}, such as condensed matter, nuclear physics, and neutron stars.
By manipulating the $s$-wave scattering length of Fermion atoms in experiments, the superfluidity phases can change smoothly from BCS to Bose-Einstein condensates (BEC).
There are numerous studies of cold Fermi gases about their static and dynamic properties~\cite{Giorgini,bloch}, demonstrating versatile advantages to explore novel superfluidity phases, quantized vortices, collective modes, and many-body effects.

The collective modes of Fermi gases can be precisely measured~\cite{bartenstein,altmeyer}, which provides a good testing ground for different aspects of many-body theories.
In particular, the breathing mode has been extensively studied~\cite{Bruun}, which is related to the equation of state and incompressibility.
In the unitary limit, the Fermi gas is characterized by infinite scattering length and its dynamic properties are insensitive to interactions.
The collective oscillation frequencies of large systems can be well described by the hydrodynamic approach~\cite{bulgac05,heiselberg} and are insensitive to effective interactions.
In contrast to large systems adopted in experiments,
the detailed quantum effects such as shell effects and superfluidity can  have large impacts for small systems.
Small systems have a finite density of states and collective modes would be different from
the continuous hydrodynamic approach. 
It is, therefore, crucial to establish a connection
 between small systems that can be described by microscopic approaches and large systems that can be described by the hydrodynamical approach.

Some of the microscopic approaches, for example, the quantum Monte Carlo method (QMC)~\cite{par-b},  are numerically very challenging for a large number of trapped particles.
The suitable microscopic method to describe the ground state of Fermi gases is the Bogoliubov de Gennes equation (BdG).
For unitary Fermi gases, the SLDA~\cite{bulgac07} is an effective superfluid DFT approach  which was developed according to experiments and QMC simulations.
The collective excitations can be in principle self-consistently described by the time-dependent BdG equation, or time-dependent density functional theory~\cite{bulgac09}.
For small amplitude collective motions, such as breathing vibrations,  QRPA (or linear response theory) can match the real-time dynamical results and is more efficient.
Several QRPA calculations have been performed for Fermi gases, and interesting insights have been achieved in such as multipole collective modes~\cite{Bruun,bruun01}, the Higgs modes~\cite{Korolyuk}, Goldstone modes~\cite{zou,Hoinka},  pairing breaking modes~\cite{Kurkjian,Hoinka} and finite-size effects~\cite{grasso}.
In addition, the transition currents in QRPA can directly reveal the dynamic mechanisms of collective modes~\cite{wang}.
In experiments, Fermi gases have usually been studied in highly elongated traps~\cite{bartenstein,altmeyer}.
By varying the trap deformations in QRPA calculations to highly prolate (elongated) shapes, we can simulate the collective modes of quasi-one-dimensional  systems.
We can also explore the collective modes of quasi-2D systems with extremely oblate (pancake) deformed traps, which are also attractive in experiments and theories~\cite{exp2d,liu2d}.
However, previous QRPA calculations of Fermi gases are not really preformed within deformed traps.
Actually, deformed QRPA has been widely applied in nuclear physics~\cite{ring} but calculations are very time consuming.

In this work, we aim to investigate the collective breathing modes of a unitary Fermi gas in very deformed traps, within  the framework of QRPA.
We can study the dynamical properties of systems in elongated and pancake traps with a small number of particles, e.g. 200 particles,  due to large computational costs.
The properties of such systems are also interesting for nuclear physics.
To see the role of finite-size effects, we also studied large systems consisting of 2000 particles.
The realistic particle numbers in experiments are about 10$^5$~\cite{bartenstein,altmeyer}, which are too large for microscopic calculations.
The calculations are performed in axial-symmetric discretized coordinate spaces based on B splines~\cite{Pei08}.
The ground states of unitary Fermi gases are described by SLDA with different parameters~\cite{bulgac07}.
The QRPA equation is based on wavefunctions from SLDA and is solved by an iterative method, known as the finite-amplitude method~\cite{fam}, instead of standard diagonalization of a huge matrix~\cite{ring}.
The used iterative solving method  is much more efficient and has been widely applied in finite quantum systems such as nuclear and chemical physics.
The results include transition strengthes as a function of oscillation frequencies, and also transition currents which can reveal the dynamic mechanisms of collective modes.

\section{\label{sec:level2}THEORETICAL FRAMEWORK}

\subsection{\label{sec:level2A}Superfluid Local Density Approximation}

For the unitary Fermi gas, the system exhibits a universal behavior governed by the densities, making it ideally suited for a DFT description~\cite{bulgac07,papenbrock}.
DFT can incorporate experimental information and Quantum Monte Carlo results within an accurate energy density functional
 and is able to describe strongly interacting systems due to the incorporation of many-body effects.
In this respect, SLDA, a superfluid extension of DFT, was developed by Bulgac and it has been applied to symmetric two-component systems~\cite{bulgac07}.
Time-dependent SLDA calculations~\cite{bulgac09} and SLDA-RPA calculations~\cite{zou} have been demonstrated to be very useful for studies of excitation properties of Fermi gases.
The SLDA has been generalized to ASLDA for spin-imbalanced systems~\cite{aslda}, which predicts the existence of Larkin-Ovchinnikov phases in elongated traps~\cite{pei10}.

We consider that the symmetric unitary Fermi gas is trapped by an axial-symmetric external potential,
$V_{\rm ext}(\rr)=\frac{1}{2} m \omega_T^2 \left(r^2+z^2/\eta^2 \right)$ , where $\eta$ denotes the elongation of the trap and $z$-axis is the principle axis. We work in trap units for which $\hbar$=m=$\omega_T$=1.
The SLDA energy density functional  is written as~\cite{bulgac07}:
\begin{equation}\label{eq2}
\begin{split}
  \epsilon(\rr) = & \alpha \frac{\tau(\rr)}{2} + \beta \frac{3(3\pi^2)^{\frac{2}{3}}\rho^{\frac{5}{3}}(\rr)}{10} - \Delta(\rr)\kappa(\rr) \\
 & + V_{\rm ext}(\rr)\rho(\rr) \textrm{,}
  \end{split}
\end{equation}
where the particle densities $\rho(\rr)$, kinetic energy density $\tau(\rr)$, and pairing density $\kappa(\rr)$, and pairing potential $\Delta(\rr)$ are
written as,
\begin{equation}\label{eq3}
\begin{split}
  &\rho(\rr) = 2\sum_{E_k<E_c}{\left|v_k(\rr)\right|^2} \textrm{,} \hspace{10pt} \tau(\rr) = 2\sum_{E_k<E_c}{\left|\nabla v_k(\rr)\right|^2} \textrm{,}\\
  &\kappa(\rr) = \sum_{E_k<E_c}{v_k^*(\rr)u_k(\rr)} \textrm{,}  \hspace{10pt} \Delta(\rr)=-g_{eff}(\rr)\kappa(\rr) \textrm{.}
\end{split}
\end{equation}
The effective pairing strength $g_{eff}(\rr)$ is obtained by renormalization according to a ultraviolet cutoff $\Lambda_c$~\cite{bulgac07}
\begin{equation}\label{eq6}
  \frac{1}{g_{eff}(\rr)} = \frac{\rho^{\frac{1}{3}}(\rr)}{\gamma} + \Lambda_c(\rr) \textrm{,}
\end{equation}
In the above equations, the SLDA parameters $\alpha$, $\beta$ and $\gamma$ are dimensionless constants.

Minimization of the energy density functional leads to a BdG set of equations:
\begin{equation}\label{eq1}
  \begin{pmatrix}
  h(\rr)-\mu & \Delta (\rr) \\ \Delta^* (\rr) & -(h^* (\rr)-\mu)
  \end{pmatrix}
  \binom{u_k(\rr)}{v_k(\rr)}
  =E_k \binom{u_k(\rr)}{v_k(\rr)} \textrm{.}
\end{equation}
where the single-particle Hamiltonian $h$ is determined by variation of SLDA energy density functional with respect to $\tau$ and $\rho$ densities.
The densities are constructed from quasi-particle wave functions $u_k(\rr)$ and $v_k(\rr)$.

Note that the SLDA form in Eq.$(\ref{eq2})$ breaks the Galilean invariance, and therefore the contribution of the current density $\vec{j}(\rr)$ should be included,
\begin{equation}\label{eq4}
\begin{split}
  \epsilon(\rr) = &\alpha \frac{\tau_c(\rr)}{2} + \beta \frac{3(3\pi^2)^{\frac{2}{3}}\rho^{\frac{5}{3}}(\rr)}{10}
  + g_{eff}{\left|\kappa_c(\rr)\right|^2} \\
  & + V_{\rm ext}(\rr)\rho(\rr) - (\alpha-1)\frac{\vec{j}^2(\rr)}{2\rho(\rr)} \textrm{,}
\end{split}
\end{equation}
where the current density $\vec{j}(\rr)$ is written as:
\begin{equation}\label{eq5}
\begin{split}
  &\vec{j}(\rr) = 2{\rm Im} \sum_{E_k<E_c} {v_k^*(\rr)\nabla v_k(\rr)} \textrm{.}
\end{split}
\end{equation}
The current density has no contribution to ground states of spin-balanced systems, but it plays a role in excited states.

The parameters $\alpha$, $\beta$ and $\gamma$ in the SLDA energy density functional can be related to physical parameters $\alpha$, $\xi$ and $\eta$~\cite{bulgac07}.
The parameter $\alpha$ is defined by the effective mass, $\alpha=m/m_{eff}$.
The parameters $\xi$ and $\eta$ are defined by the particle-number density $\rho=N/V=\frac{k_F^3}{3\pi^2}$, the chemical potential $\mu=\xi\epsilon_F$ ($\xi$ is called the Bertsch parameter), and the pairing gap $\Delta=\eta\epsilon_F$ of the homogeneous Fermi gas. These physical quantities can be determined by QMC calculations~\cite{par-b,gezerlia} and experimental measurements~\cite{navon}.
There are several sets of parameters $\alpha$, $\beta$ and $\gamma$ adopted in the literature, as shown in Table \ref{tab1}.

\renewcommand\arraystretch{1.5}
\begin{table}[b]
\begin{tabular}{p{1.3cm}<{\centering}p{1.3cm}<{\centering}p{1.3cm}<{\centering}p{1.3cm}<{\centering}p{1.3cm}<{\centering}p{1.3cm}<{\centering}}
  \hline
  \hline
  Parameter sets & a~\cite{par-a} & b~\cite{par-b} & c~\cite{par-c} & d~\cite{zou} & e~\cite{par-e} \\
  \hline
  $\alpha$ & 1.14 & 1.12 & 1.104 & 1.00 & 0.812 \\
  $\beta$ & -0.553 & -0.520 & -0.417 & -0.430 & -0.712 \\
  1/$\gamma$ & -0.0906 & -0.0955 & -0.0347 & -0.0767 & -0.0705 \\
  $\xi$ & 0.422 & 0.440 & 0.374 & 0.376 & 0.449 \\
  $\eta$ & 0.504 & 0.486 & 0.651 & 0.500 & 0.442 \\
  \hline
  \hline
\end{tabular}
  \caption{The five sets of SLDA parameters adopted in this work. }
  \label{tab1}
\end{table}

\subsection{\label{sec:level2B}Finite Amplitude Method for Quasi-Particle Random-Phase Approximation}

The standard QRPA equation includes many particle-hole excitations~\cite{ring},
and the resulted configuration space is huge for deformed cases and continuum. For example,  with 1000 single-particle levels,  the matrix which needs to be diagonalized
would have a dimension of the order of $\thicksim$10$^6\times$10$^6$.
On the other hand, the QRPA equation can be derived from the time-dependent Hartree-Fock-Bogoliubov equation by assuming that
the collective motion has small amplitudes~\cite{fam}. Here we refer to the finite-amplitude method for QRPA (FAM-QRPA), which 
allows to solve QRPA iteratively and much more efficiently, by avoiding the calculation of a huge configuration matrix.
The FAM-QRPA has been applied to  multipole giant resonances and pygmy resonances in nuclei~\cite{fam1,fam,wang,markus}.
Our solver is implemented in the cylindrical coordinate space which is useful for descriptions of very deformed and weakly bound systems.

In FAM-QRPA, different collective modes of systems are obtained by responses to various external oscillating fields.
We solve the non-linear FAM-QRPA equations as follows~\cite{fam},
\begin{equation}\label{eq7}
\begin{split}
  (E_\mu+E_\nu-\omega)X_{\mu\nu}(\omega) + \delta H_{\mu\nu}^{20}(\omega) = -F_{\mu\nu}^{20}(\omega) \textrm{,} \\
  (E_\mu+E_\nu+\omega)Y_{\mu\nu}(\omega) + \delta H_{\mu\nu}^{02}(\omega) = -F_{\mu\nu}^{02}(\omega) \textrm{,}
  \end{split}
\end{equation}
where $ \delta H_{\mu\nu}^{20}(\omega)$ ,$ \delta H_{\mu\nu}^{02}(\omega)$ are the induced Hamiltonian.
$ F_{\mu\nu}^{20}$, $ F_{\mu\nu}^{02}$ correspond to the external field, for which $F=r^{L}Y_{LK}\left(\theta,\varphi \right)$ is the $L^{th}$-order operator for multipole modes.
For the monopole mode (breathing mode), $F$ takes $F=r^2 Y_{00}\left(\theta,\varphi \right)$.
We solve the transition amplitudes $X(\omega)$ and $Y(\omega)$ iteratively at each collective excitation frequency $\omega$.
To smooth the transition strength, an imaginary part is included in the frequency as $\omega+i\gamma$. The non-linear FAM-QRPA equation
is solved by using the modified Broyden method~\cite{byd}.

The induced fields can be obtained by linear functions of induced densities.
For SLDA,  the induced mean-field potential $\delta U$, induced pairing potential $ \delta\Delta$
and the induced time-odd field $\delta h_{odd}$ can be written as,
\begin{eqnarray}
 \delta U  &=&  \beta\frac{({3\pi^2})^{\frac{2}{3}}}{3\rho^{\frac{1}{3}}}\delta\rho - (\frac{g_{eff}}{3\gamma\rho^{\frac{2}{3}}})^2 (\kappa^*\Delta+\Delta^*\kappa)\delta\rho \nonumber \\
           &+& \frac{g_{eff}}{3\gamma\rho^{\frac{2}{3}}}(\Delta\delta\kappa^{-}+\Delta^*\delta\kappa^{+}) + \frac{2\left|\Delta\right|^2}{9\gamma\rho^{\frac{5}{3}}}\delta\rho \textrm{,} \label{eq18} \\
 \delta\Delta^{\pm}  &=& \frac{g_{eff} \Delta}{3\gamma\rho^\frac{2}{3}}\delta\rho - g_{eff}\delta\kappa^{\pm} \textrm{,} \label{eq19} \\
 \delta h_{\rm odd}  &=& \frac{i}{2\rho}(\alpha-1)(\nabla\cdot\delta\vec{j}+\delta\vec{j}\cdot\nabla) \textrm{,} \label{eq20}
\end{eqnarray}
The induced fields are based on transition density $\delta\rho$, transition pairing density $\delta\kappa$ and transition current $\delta\vec{j}$.
These transition densities are constructed from the set of quasiparticle wavefunctions $U$, $V$ and transition amplitudes $X$, $Y$.
The various transition densities are given as,
\begin{eqnarray}
  \delta\rho       &=& UXV^T+V^*Y^TU^{\dag} \textrm{,}\label{eq21}\\
  \delta\kappa^{+} &=& UXU^T+V^*Y^TV^{\dag} \textrm{,}\label{eq22}\\
  \delta\kappa^{-} &=& V^{*}X^{\dag}V^{\dag}+UY^{*}U^{T} \textrm{,}\label{eq22}\\
  \delta\vec{j}    &=& \frac{1}{2i}\left[ (\nabla U^*)YV^* + UX(\nabla V)  \right.\nonumber\\
                   &-& \left. (\nabla U)XV - U^*Y(\nabla V^*) \right] \textrm{.}\label{eq23}
\end{eqnarray}
Finally the total transition strength function $S(F;\omega)$ of collective modes can be obtained
from the converged transition amplitudes $(X,Y)$ from  FAM-QRPA solutions as
\begin{equation}\label{eq24}
  S(F;\omega) = -\frac{1}{\pi} \textrm{Im}\sum_{\mu<\nu}\left[ F_{\mu\nu}^{20*}X_{\mu\nu}(\omega) + F_{\mu\nu}^{02*}Y_{\mu\nu}(\omega) \right] \textrm{.}
\end{equation}

\begin{figure}
  \includegraphics[width=0.35\textwidth]{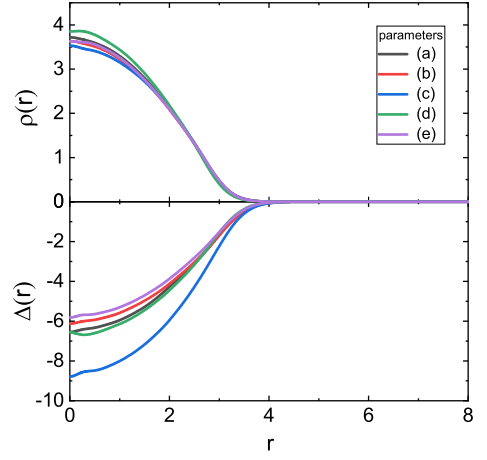}\\
  \caption{(Color online) The ground-state particle densities $\rho(r)$ and pairing gaps $\Delta(r)$ with 200 particles in a spherical trap. Calculations are performed with different SLDA parameters as given in Table \ref{tab1}. }
  \label{fig1}
\end{figure}

\begin{figure}[t]
  \includegraphics[width=0.4\textwidth]{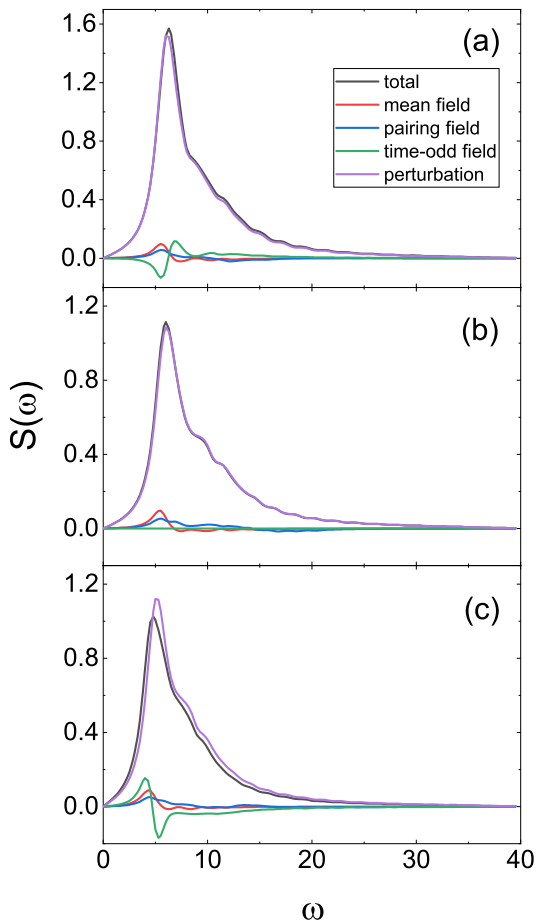}\\
  \caption{(Color online) The total response strengthes and  contributions of induced fields given by SLDA-QRPA with 200 particles in a spherical trap with different SLDA parameter sets (a), (d) and (e), respectively,  as listed in Table \ref{tab1}. }
  \label{fig2}
\end{figure}

\begin{figure}[t]
  \includegraphics[width=0.4\textwidth]{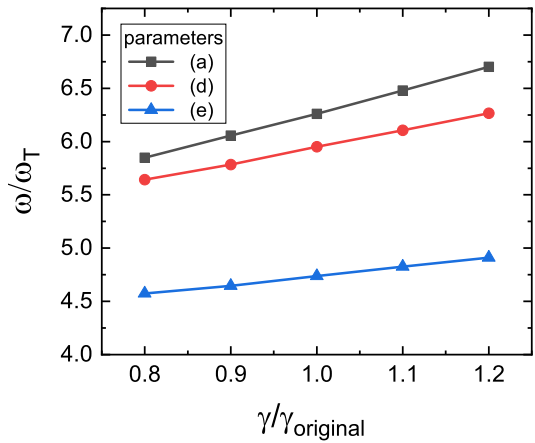}\\
  \caption{(Color online) The monopole oscillation frequencies of 200 particles in a spherical trap with varying pairing parameters $1/\gamma$.
  Calculations are performed with SLDA parameter sets (a), (d), and (e), as listed in Table~\ref{tab1}. }
  \label{fig3}
\end{figure}

\section{\label{sec:level3}Results and Discussions}

\subsection{\label{sec:level3A} Unitary Fermi gases in spherical traps }

We first consider a unitary Fermi gas of 200 particles within a spherical harmonic trap. The numerical accuracy depends on the sizes of coordinate spaces and mesh spacing.
The mesh spacing is taken as 0.2 in this work. The adopted energy cutoff is about 20 for renormalization.
Firstly, the ground states of the systems in a spherical trap are calculated.
The ground state properties are relevant for collective oscillation frequencies.
Fig.\ref{fig1} shows that number density and  pairing potential with different SLDA parameters as shown in Table.\ref{tab1}.
We see the densities are more or less similar for different parameters.
However, the parameter set (c) resulted in a particular large pairing gap which has the largest pairing strength of $1/\gamma$=$-$0.0347.

The calculated monopole excitation peak frequencies $\omega$ with different SLDA parameters are give in Table.\ref{tab2}.
The peak frequency can also be estimated by $\omega=m_1/m_0$, where  the energy-weighted sum rule is defined as $m_k = \int{d\omega S(F;\omega)\omega^k}$~\cite{bertulani}.
We see larger discrepancies in  peak-$\omega$ values for the used different parameter sets.
The largest frequency $\omega$ is obtained with parameter-(c) in Table \ref{tab1}, which corresponds to
 the largest pairing strength.
In Table \ref{tab2}, the differences between $m_1/m_0$ and $\omega$  from peaks are rather small.
The lowest $\omega$ is obtained from parameter-(e) corresponding to the smallest $\alpha$=0.0812 and the largest effective mass.
In analogy to nuclear monopole resonances, the breathing mode peak frequencies are inversely proportional to $\sqrt{m^{*}}$~\cite{Nasc09}.
In addition, the monopole frequencies are also inversely proportional to the rms radii~\cite{bertulani}.
Small systems have small radii and thus the frequencies are very large.
The obtained $\omega$ values are much larger than the hydrodynamical results of $\omega=2$ for spherical systems~\cite{bulgac05,heiselberg},
duo to the finite size effects.
For large systems, it is known that QRPA calculations of 10$^4$ particles in a spherical trap can match hydrodynamical results~\cite{grasso}.
For small systems, we see that quantum effects play a significant role and
it is a good testing ground for different parameters.

\renewcommand\arraystretch{1.5}
\begin{table}[b]
\begin{tabular}{p{2.5cm}<{\centering}p{1.05cm}<{\centering}p{1.05cm}<{\centering}p{1.05cm}<{\centering}p{1.05cm}<{\centering}p{1.05cm}<{\centering}}
  \hline
  \hline
  Systems & a & b & c & d & e \\
  \hline
  $\omega$(FAM-QRPA) & 6.261 & 6.033 & 6.943 & 5.951 & 4.738 \\
  $m_1/m_0$ & 6.250 & 6.054 & 7.021 & 6.008 & 4.804 \\
  $E_{gs} $ & 840.6 & 859.8 & 795.4 & 795.0 & 857.5 \\
  $\sqrt{<r^2>}$ & 2.15 & 2.17& 2.17 & 2.11 & 2.16 \\
  \hline
  \hline
\end{tabular}
  \caption{The monopole oscillation peak frequencies of 200 particles in a spherical trap, calculated with SLDA-QRPA with different parameters.
  The peak energies are compared with the sum rule results $m_1/m_0$. The ground state energies and rms radii are also given.
 }
  \label{tab2}
\end{table}

To further analysis the role of different SLDA parameters, we display
the different contributions to transition strengthes related to, i.e., the mean field, the pairing field, and the time-odd field,  to the total the transition strengthes,
as shown in Fig.\ref{fig2}.
In Fig.\ref{fig2}, transition strengthes corresponding to three parameter sets (a), (d) and (e) are shown.
Note that the calculated transition strengthes are given with a smoothing parameter $\gamma$ of 0.2.
The main contribution to the total response strength is from the external perturbation field, which is determined by ground state properties.
The contribution of induced mean field has a lower frequency than the main peak frequency.
The contribution of induced pairing fields is not significant.
The induced time-odd field, based on the transition current $\delta\vec{j}$, plays a significant role when the effective mass is not equal 1.
For $\alpha$ smaller than 1 in parameter-(e), the $\delta\vec{j}$ field has a positive contribution in low-energy
and thus shifts the frequency remarkably to low energies (see Table \ref{tab2}).
In contrast, for $\alpha$ larger than 1 in parameter-(a), the $\delta\vec{j}$ field has a oppositive contribution and
shifts the frequency to high energies. This is consistent with the role of effective mass, in which a large effective mass (or a small $\alpha$) results in a small monopole frequency.
We demonstrated that the induced time-odd term and the effective mass have significant consequences in collective modes of small systems.

The role of pairing in the equation of state and collective modes is very interesting.
For example, it is interesting to know the differences between superfluid and nonsuperfluid hydrodynamics~\cite{wright}.
The monopole frequencies with varying pairing interaction strengthes are given in Fig.\ref{fig3}.
Note that the influences of pairing include the static pairing and included pairing.
We see that with the parameter (a), the monopole frequency
increased linearly by 7$\%$ when pairing strength increased by $20\%$. The dependence on pairing strength is less prominent for cases with smaller excitation frequencies.
This explains that parameter (c) with the largest pairing strength (about 2.6 times larger than (a)) corresponds  to the largest monopole frequency, compared
to other parameter sets.
The monopole frequencies $\omega$ are less affected even if we turn off the induced pairing interactions,
implying the static pairing rather than the induced pairing is dominated. In deeded, the largest monopole peak frequency $\omega$ is related to the largest ground-state pairing field $\Delta$ at the trap center as shown in Fig.\ref{fig1}.
In the cold Fermi gases close to unitary, the pairing plays a role to increase the collective oscillation frequency~\cite{grasso}.
In contrast, the pairing can increases collectivity and shifts down nuclear giant monopole resonance energies~\cite{bertulani,khan}.
On the other hand, the pairing can increase frequencies of nuclear pygmy resonances associated with dilute halo surfaces~\cite{matsuo}, which is similar to unitary Fermi gases.
Note that calculations in this work are performed at zero temperature, and a finite temperature in experiments can significantly reduce the pairing.
We see that the monopole frequency of a small spherical system can be very much dependent on SLDA parameters. We will return to this observation in the later part.

\begin{figure}[t]
  \includegraphics[width=0.4\textwidth]{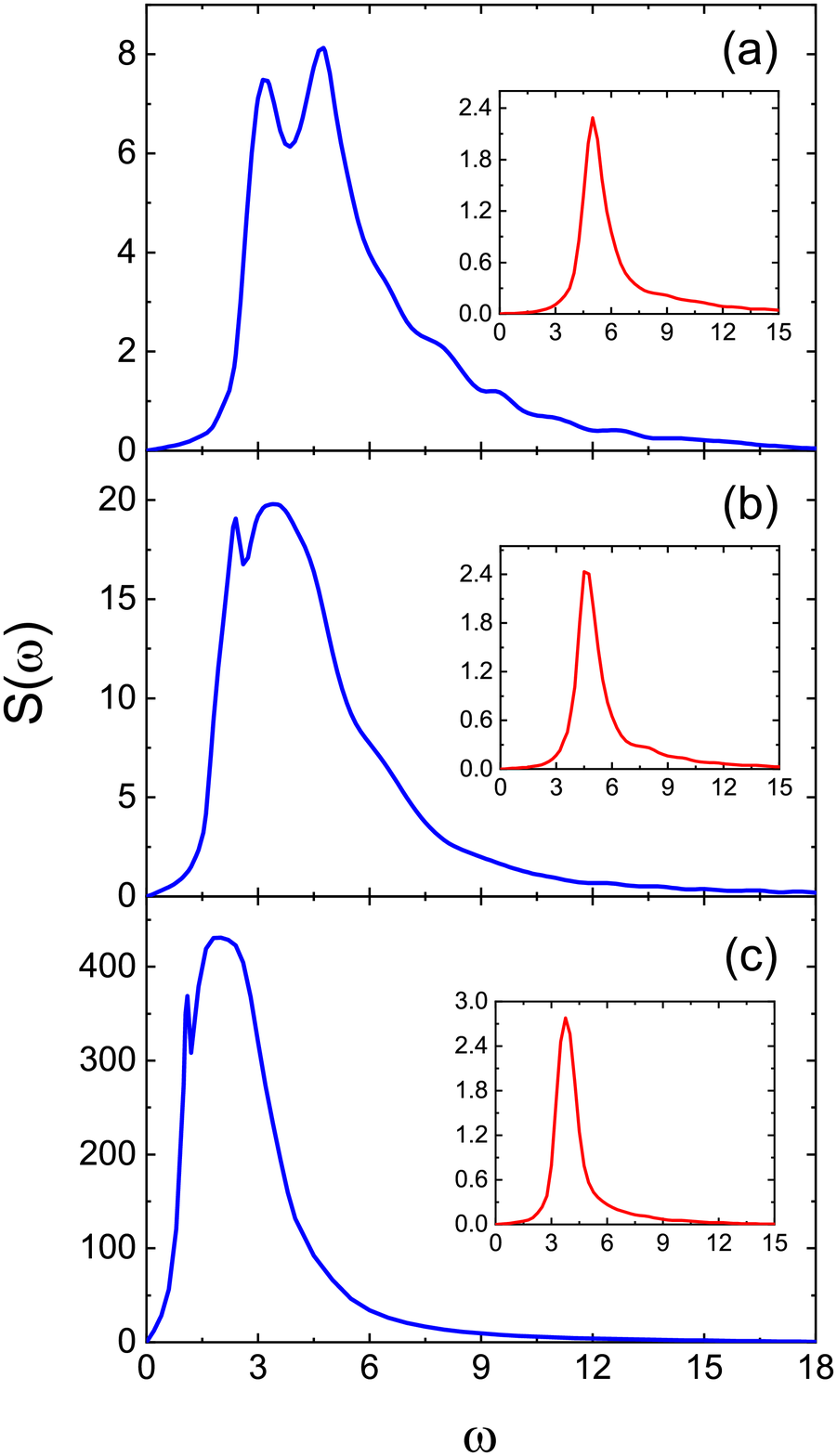}\\
  \caption{(Color online) The transition strengthes of monopole mode of 200 particles in prolate traps with trap aspect ratios of $\eta$=3 (a), $\eta$=5 (b), $\eta$=20 (c), respectively.
  The inset figures show the corresponding transition strengthes of radial modes. }
  \label{fig4}
\end{figure}

\begin{figure}[t]
  \includegraphics[width=0.4\textwidth]{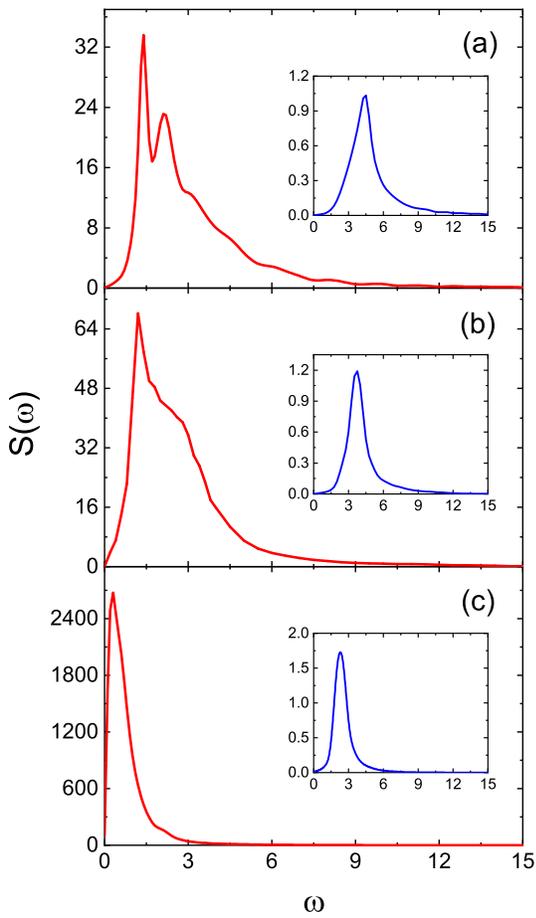}\\
  \caption{(Color online) The transition strengthes of monopole mode of 200 particles in oblate traps with trap aspect ratios of $\eta$=3 (a), $\eta$=5 (b), $\eta$=20 (c), respectively.
  The inset figures show the corresponding transition strengthes of axial modes.}
  \label{fig5}
\end{figure}


\subsection{\label{sec:level3D} Collective frequencies of deformed systems}

We have studied monopole modes of a unitary Fermi gas of 200 particles in prolate (elongated shape) traps  $V_{ext}(\rr)=\frac{1}{2}m\omega_T^2(r^2+z^2/\eta^2)$ and oblate (pancake shape) traps $V_{ext}(\rr)=\frac{1}{2}m\omega_T^2(r^2/\eta^2+z^2)$.
The trap aspect ratio is defined by $\eta$.
The collective transition strengthes of prolate and oblate traps with the parameters-(a) are shown in Fig.\ref{fig4} and Fig.\ref{fig5}, respectively.
In Fig.\ref{fig4},  the monopole transition strength with $\eta$=3 has two peaks at frequencies of 3.21 and 4.71, respectively.
With $\eta$=5, the first peak is shifted to a lower energy of 2.42 and the second peak becomes prominent with an energy of 3.47.
With increasing $\eta$, the second peak becomes dominated and the first peak becomes narrower.
The calculated radial modes are also displayed as insets in Fig.\ref{fig4}. It is known that
the axial modes and radial modes are separated in well-deformed cases~\cite{wang}. As shown, the radial modes
have a single peak which slowly shifts to low energies as deformation increases.
The widths of the first axial mode and the radial mode become narrower with increasing deformations. 
The resonance widths can be related to the damping widths in experiments.
 In elongated traps, however, the transition strengthes
of radial modes are too small compared to axial modes.
The appearance of the second mode in elongated traps is rather unusual and we will discuss it in following sections.

In oblate traps as shown in Fig.\ref{fig5}, the modes correspond to a small deformation $\eta$=3 also have two peaks.
With increasing $\eta$, the main peak shifts to lower energy and becomes narrower. At $\eta$=20, the second
peak disappears and the breathing mode is a single narrow resonance.
The axial modes are suppressed in pancake traps. The axial modes also shift to lower energies and become narrower, approaching the hydrodynamic limit.
The widths of radial and axial modes in pancake traps are comparable.

In both elongated and pancake traps, the obtained peak frequencies are much higher
than indicated by the quasi-1D ($\omega=\sqrt{12/5}\omega_z$) and quasi-2D ($\omega=\sqrt{3}\omega_r$)  hydrodynamical results~\cite{heiselberg}.
The deviation is particularly significant in elongated traps.
 The peak frequencies actually decrease slowly with increasing prolate deformations, while the deformation scaling is more reasonable in oblate traps.
In contrast, the radial modes in elongated traps and axial modes in pancake traps are approaching the hydrodynamical results of  quasi-1D ($\omega=\sqrt{10/3}\omega_r$) and quasi-2D ($\omega=\sqrt{8/3}\omega_z$) systems~\cite{heiselberg}
as trap deformation increases.
The differences in collective frequencies due to different parameters also decreased at large deformations.

\begin{figure}[t]
\flushleft
  \includegraphics[width=0.48\textwidth]{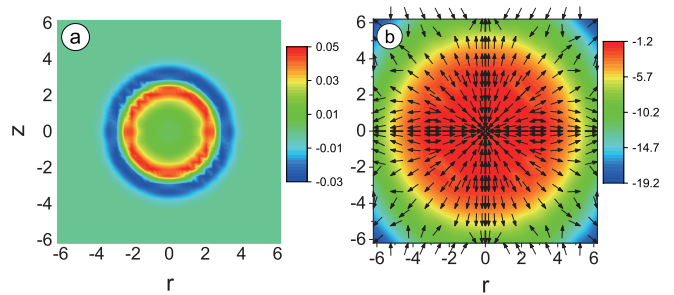}
  \caption{(Color online) (a) Calculated induced transition density $\delta\rho(r,z)$, and (b) induced current density  $\delta \vec{j}(r,z)$  of monopole modes of a spherical system
  of 200 particles. }
  \label{fig6}
\end{figure}

\begin{figure}[t]
\flushleft
  \includegraphics[width=0.48\textwidth]{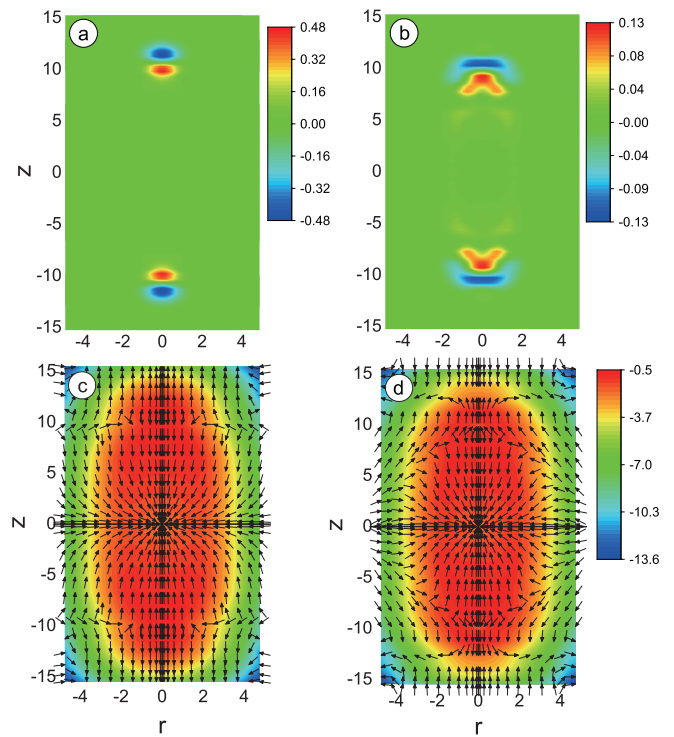}
  \caption{(Color online) (a) For systems with 200 particles within elongated trap of $\eta$=5,   calculated induced transition density $\delta\rho(r,z)$: (a) for peak at 2.42 and (b) for peak at 3.47.
    Calculated induced current density $\delta \vec{j}(r,z)$: (c) for peak at 2.42 and (d) for peak at 3.47.  }
  \label{fig7}
\end{figure}

\begin{figure}[t]
\flushleft
  \includegraphics[width=0.48\textwidth]{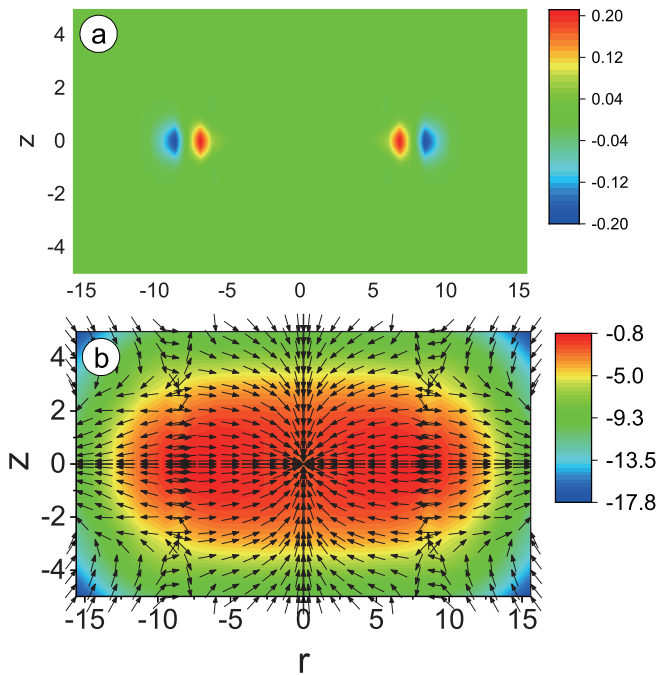} \\
  \caption{(Color online) (a) Calculated induced transition density, (b) induced current density of a system of 200 particles in a pancake trap with $\eta$=5. The corresponding
  energy of the monopole mode is 1.24.  }
  \label{fig8}
\end{figure}

\subsection{\label{sec:level3E} Transition densities and currents}

In QRPA calculations, the obtained transition densities $\delta\rho$ and transition currents $\delta \vec{j}$ can
directly reveal the dynamical mechanism of collective modes.
An additional interesting ingredient is provided by the superfluidity and its effects on the flow pattern. 
The transition densities and currents are also major ingredients in the hydrodynamical approach.
It is possible that the transition densities and currents can be obtained by analysis of angle-resolved scattering experiments.
It is known that the flow patterns of collective modes are dependent on excitation energies and deformations~\cite{wang}.

To study the mechanisms of collective modes in different traps, we calculated the transition
densities and currents with the SLDA parameter of $\alpha=1.14$, $\beta=-0.553$, and $1/{\gamma}=-0.0906$.
The results of the spherical trap are shown in Fig.\ref{fig6}.  For the transition density,
a ring structure appears with a radius of 3, indicating that the breathing mode is a surface mode.
The transition density illustrates clearly the breathing behavior with oppositive values between inside and outside rings.
The integral of the total transition density is close to zero.
For the transition current, the compressional flow is inward within the ring, but the flow is outward beyond the ring.
This is because the velocity changes direction at outer surface, where the density is about 12$\%$ of the center density. 
Note that the same time-evolution forms of oscillating transition densities and currents are assumed. 
We illustrated that the compressional flow in spherical cases has a simple pattern.

For deformed systems, we show the transition densities and currents in the elongated traps with $\eta$=5 in Fig.\ref{fig7}.
There are two resonance peaks at $\omega$=2.42 and $\omega$=3.47.
From the transition densities, we can see that both modes mainly related to vibrations at axial ends of the elongated system.
indicating both modes are axial modes.
From the transition currents, we can see the lowest modes have a clear compressional flow pattern towards the center.
The flow pattern has 3 nodes in the axial direction.
For the second peak, the transition current is more complex and its transition density involves a larger spatial region, associated with
a higher excitation energy.
Actually, the current of the second peak has 5 nodes in the axial direction, which is distinct to the first peak with 3 nodes.   Naturally the current patterns
become complex as excitation energies increase, in analogy to quantum topology excitations.

The transition density and currents of the oblate $\eta$=5 trap are shown in Fig.\ref{fig8}.
The transition density shows that the vibrations are mainly at the surrounding edge of the pancake system.
The transition currents indicate that it is a simple radial mode. Close to the center, the flow is inwards and  the current has 3 nodes in the radial axis.
Actually the flow of the pancake system is similar to the flow of the first peak in the elongated system when rotated by 90 degrees.
The transition densities show that collective modes in elongated and pancake systems are axial and radial modes, respectively.
However, the transition currents show that the collective modes are nontrivial as the presence of several nodes indicates.
A full 3D calculations should be adopted for more realistic studies of flow patterns.

\begin{figure}[t]
\flushleft
  \includegraphics[width=0.48\textwidth]{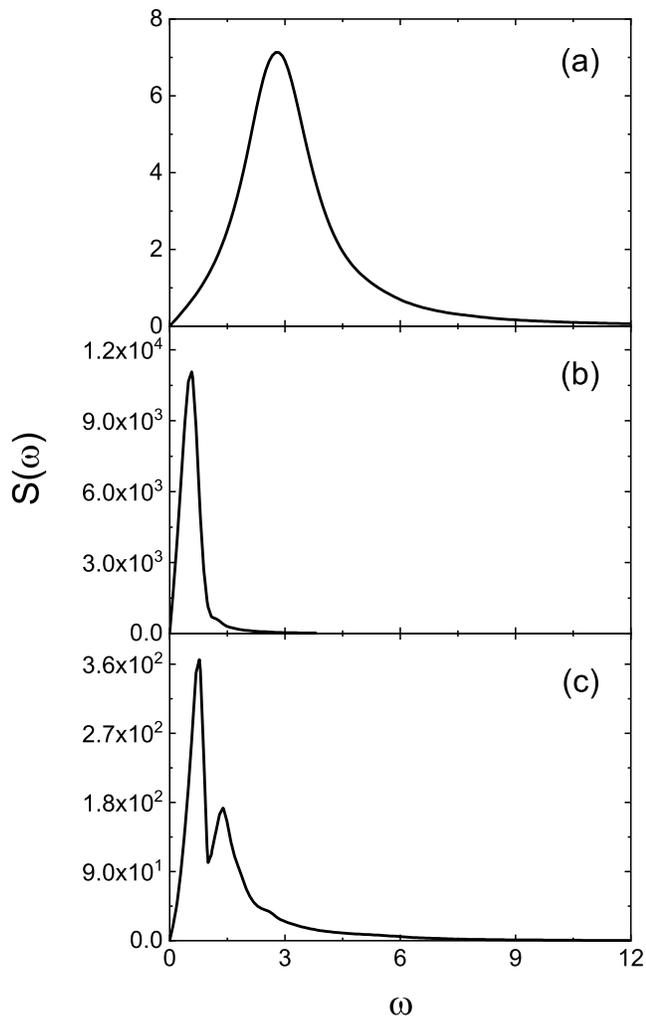} \\
  \caption{(Color online) Calculated transition strengthes with 200 particles, (a) in a spherical trap with $1/{\gamma}$=-0.4,   (b) in an elongated trap of $\eta$=20 with parameter $1/{\gamma}$=-0.4,
  (c) in an elongated trap of $\eta$=20 with parameter $1/{\gamma}$=-0.2.
   }
  \label{fig9}
\end{figure}

\begin{figure}[t]
\flushleft
  \includegraphics[width=0.48\textwidth]{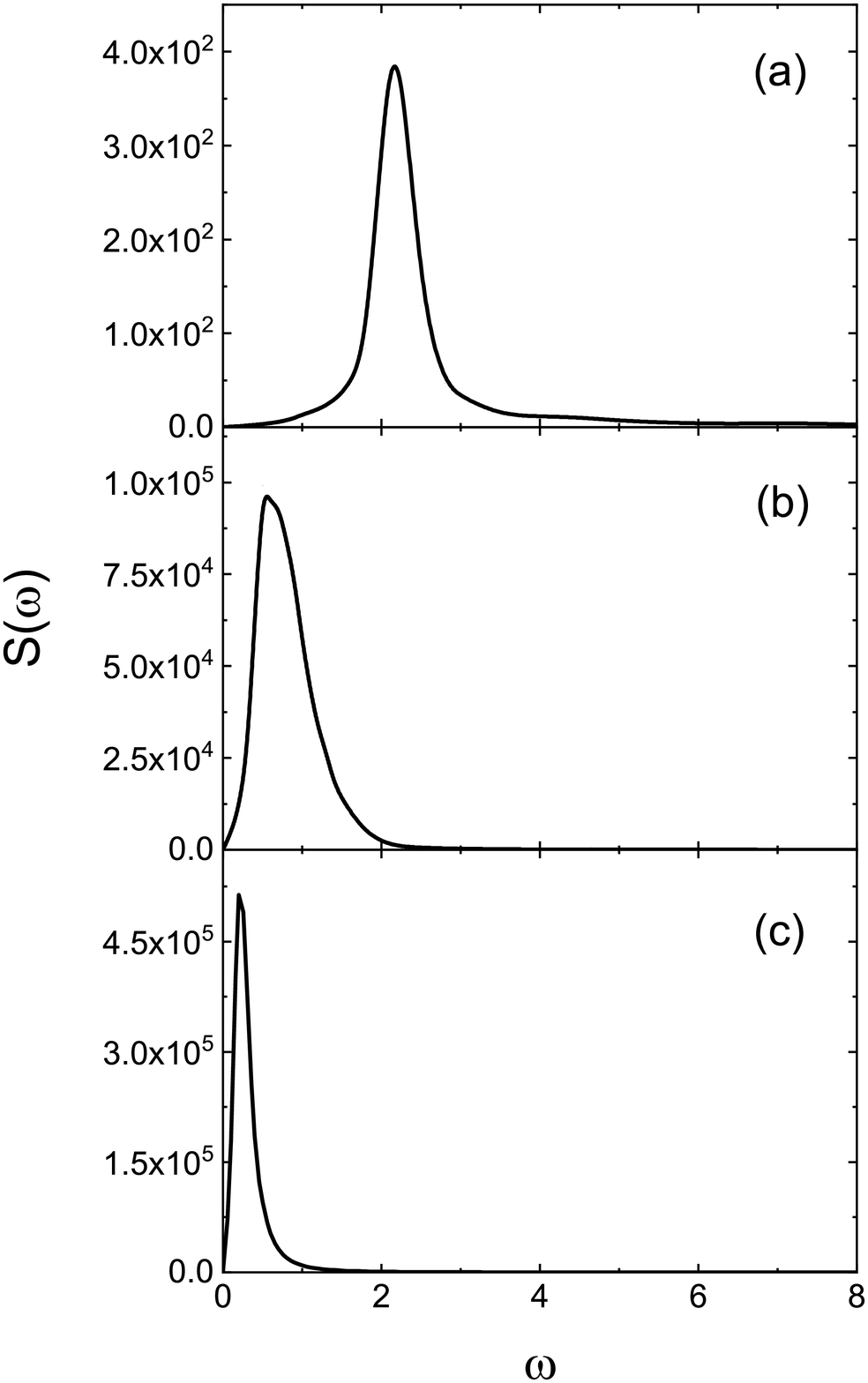} \\
  \caption{(Color online)
Calculated transition strengthes with 2000 particles with $1/{\gamma}$=-0.4, (a) in a spherical trap,   (b) in an elongated trap of $\eta$=20,
  (c) in a pancake trap  of $\eta$=20.}
  \label{fig10}
\end{figure}

\subsection{\label{sec:level4} Towards large systems}

It is interesting to study the evolution of collective modes from small to large systems.
However, it is a very demanding task for SLDA and QRPA calculations of large systems in deformed traps.
We performed calculations of 2000 particles in deformed traps, to compare with systems of 200 particles.
It has been shown that QRPA calculations of 10$^4$ particles can match the hydrodynamical approach~\cite{grasso}.
In Ref.\cite{grasso},  calculations are performed based on the mean-field approximation in spherical traps at the BCS side of the crossover.
However, with the parameters in Table \ref{tab1}, our SLDA-QRPA calculations of 2000 particles in the spherical trap give an even higher excitation energy than that of 200 particles.
This is obviously inconsistent with the finite-size scaling and is mainly because the SLDA pairing strength is too large compared
to Ref.\cite{grasso}.  To simulate the mean-field approach, in which the pairing strength $g=4\pi \hbar^2 a_s/m$~\cite{grasso} and $a_s$ is the $s$-wave scattering length,
we adopt a SLDA parameter set as $\alpha$=1, $\beta$=$-$0.509, $1/\gamma$=$-$0.41, corresponding to strong coupling systems with $k_Fa_s$=$-$0.6.
In this case, the pairing strength is significantly decreased compared to the parameter sets in Table \ref{tab1}.
Actually, the finite-size dependence is more significant in the pairing gap than in the Bertsch parameter $\xi$~\cite{Braun}. 

In Fig.\ref{fig9}, the calculated monopole modes of 200 particles with the reduced pairing are shown.
Panel (a) shows the transition strength in a spherical trap with $1/\gamma$=$-$0.401.
The monopole peak energy is 2.78, which is much lower than the results shown in Table \ref{tab2}.
We demonstrated again that the monopole mode is shifted to lower energy with reduced pairing.
Panel (b) in Fig.\ref{fig9} shows similar calculations in elongated trap with $\eta$=20.
The monopole mode is now a single peak with the peak energy of 0.53.
Lastly, Panel (c) shows similar calculations in elongated trap with $\eta$=20 but with with $1/\gamma$=$-$0.2.
In this case, the monopole mode has a lowest peak at 0.74 and a less prominent second peak at 1.35, respectively. 
 This is consistent with Fig.\ref{fig4}, in which the second peak is prominent due to large pairing strength.
 This indeed shows that 
the appearance of the second monopole peak in elongated traps is merely due to large pairing strength.

Fig.\ref{fig10} shows the results of 2000 particles with the reduced pairing of $1/\gamma$=$-$0.401.
Panel (a) in Fig.\ref{fig10}  shows the results in a spherical trap.
The obtained monopole energy is 2.11, which is very close to the hydrodynamical result of 2.0~\cite{heiselberg}.
In addition, the monopole resonance becomes narrower towards larger systems. 
Based on Fig.\ref{fig9}(a) and Fig.\ref{fig10}(a), the evolution of the peak energy from 2.78 of 200 particles to 2.11 of 2000 particles
shows very reasonable finite-size effect in spherical systems. 
Fig.\ref{fig10}(b) and Fig.\ref{fig10}(c) show the results of 2000 particles in elongated and pancake traps with $\eta$=20,   respectively.
The resulting peak energies are 0.42 and 0.19, respectively. The obtained breathing modes in pancake traps have extremely small damping widths, compared to spherical and elongated traps.
 The results are still higher than the hydrodynamical results considering the trap deformations. This is particularly serious for elongated systems.
The results of monopole energies imply that the effective pairing interaction should be much reduced in quasi-1D and quasi-2D systems compared to 3D systems.
Indeed, the coupling strength in 1D systems is defined by $g_{1D}\approx2\hbar^2 a_s/m {a_{r}}^2$~\cite{Bergeman,xjliu,bloch}, where $a_{r}$ denotes the transverse oscillator length.
This is very different from the effective 3D coupling $g_{3D}=4\pi \hbar^2 {a_{s}/m}$~\cite{grasso,bloch}. The 2D scattering length can also be related to the 3D scattering length through
renormalization of the bound state energy~\cite{orel,liu2d,bloch}.
The current SLDA parameters are developed for 3D systems and are mostly likely can handle only modest deformation~\cite{bulgac07}.
Therefore, for very elongated and pancake systems, the SLDA functional may need to be improved by including gradient terms or by adjusting effective parameters, but
further quantitative studies are beyond the scope of this work.

\section{\label{sec:level4}Conclusions}

In summary, we implemented the SLDA-QRPA approach in deformed traps and studied
the collective breathing modes of unitary Fermi gases, in the context that collective modes can be precisely measured by experiments. 
For a small system in a spherical trap, very different monopole frequencies are obtained with different SLDA parameters,
providing a good testing ground for effective parameters.
The strong pairing can increase the collective monopole frequency. The larger effective mass leads to a lower frequency.
The role of the time-odd current $\vec{j}$ term is consistent with the role of effective mass.

From spherical to elongated traps, there are two resonance peaks and the appearance of the second peak is merely due to the strong pairing.
From spherical to pancake traps, the single peak becomes narrow and is shifted to a lower-energy radial mode.
The deformation scaling is reasonable in pancake systems while collective frequencies in elongated systems are much higher than hydrodynamical results.
This illustrated the different behaviors between transitions from 3D to either a quasi-1D system or  to a quasi-2D system.
The peak frequencies of small systems of 200 particles are much higher than the hydrodynamical approach due to finite-size effects.
Towards large systems, the monopole frequency of 2000 particles in a spherical trap is 2.1 which is very close the the hydrodynamical result of 2.0.
The resonance widths, i.e., the damping,  are also reduced towards large systems. 
We also investigated  transition densities and currents, which show that the collective mechanisms are nontrivial.
The flow pattern of the second peak in elongated traps has 5 nodes along the z-axis, while the first peak has 3 nodes.

The results of monopole frequencies of extremely deformed systems indicate that the pairing interaction strength in SLDA should be reduced in low-dimensional systems
to reproduce the hydrodynamical limit.
In this sense, the experimental studies of small and medium systems are very anticipated. This will be useful for understandings of finite-size effects,
 trap deformation effects, and the role of effective interactions. This would also have  strong interdisciplinary interests for
nuclear physics and other finite quantum systems.
Our framework will be helpful for future studies of large systems and various collective modes, including pairing modes.

\begin{acknowledgments}
We thank useful comments by A.Bulgac.
 This work was supported by  National Key R$\&$D Program of China (Contract No. 2018YFA0404403),
 and the National Natural Science Foundation of China under Grants No.11835001,11790325.
 This work was also partially supported (M.K.) by the Academy of Finland under the Academy project no. 318043.
We also acknowledge that computations in this work were performed in Tianhe-1A
located in Tianjin and Tianhe-2
located in Guangzhou.
\end{acknowledgments}

\nocite{*}


\end{document}